\documentclass[preprint,aps,showpacs,nofootinbib]{revtex4}

\usepackage{epsfig,amssymb,amsmath}

\def\bea{\begin{eqnarray}}
\def\eea{\end{eqnarray}}
\def\bec{\begin{center}}
\def\ec{\end{center}}

\def\beq{\begin{equation}}
\def\eeq{\end{equation}}

\begin{document}
\draft
\tighten
\title{\large \bf SUSY Breaking at the Tip of Throat and Mirage Mediation\footnote{Talk given at From Strings To LHC
Workshop, Jan 2007, Goa, India}}
\author{
Kiwoon Choi\footnote{email: kchoi@hep.kaist.ac.kr}}
\address{
Department of Physics\\ Korea Advanced Institute of Science and
Technology\\ Daejeon 305-701, Korea}
\date{\today}

\vspace{1cm}

\begin{abstract} We discuss some features of supersymmetry breaking
induced by a brane-localized source which is stabilized at the IR
end of warped throat, and also the resulting mirage mediation
pattern of soft terms of the visible  fields which are localized
in the bulk space corresponding to
the UV end  of throat. Such supersymmetry breaking scheme  can be
naturally realized  in KKLT-type string compactification, and
predicts highly distinctive pattern of low energy superparticle
masses which might be tested at the LHC.
\end{abstract}
 \maketitle

\section{introduction}

Low energy supersymmetry (SUSY) \cite{nilles} is one of the prime
candidates for physics beyond the standard model at the TeV scale
which will be probed soon at the LHC. One key question on low energy
SUSY is the origin of soft SUSY breaking terms of the visible
gauge/matter superfields in low energy effective lagrangian. Most
phenomenological aspects of low energy SUSY are determined by those
soft terms which would be induced by the auxiliary components of
some messenger superfields \cite{kane}. To identify the dominant
source of soft terms and determine low energy superparticle masses,
one needs to compute the relative ratios between the auxiliary
components of different messenger fields. This requires an
understanding of how the messenger fields are stabilized at a
phenomenologically viable vacuum.

In string theory, moduli fields (including the string dilaton) are
plausible candidates for the messenger of SUSY breaking
\cite{munoz}. In addition to  moduli fields, the 4-dimensional
supergravity (SUGRA) multiplet provides a model-independent source
of SUSY breaking called anomaly mediation \cite{anomaly}, which is
most conveniently described by the 4D SUGRA compensator. Recent
KKLT construction \cite{kklt} of de Sitter (dS) vacuum possibly
stabilizing all moduli in Type IIB string theory has led to a new
pattern of soft terms named ``{\it mirage mediation}"
\cite{mirage1,mirage2}. In KKLT compactification, 4D $N=1$  SUSY
is broken by anti-brane (or any kind of brane providing
SUSY-breaking dynamics) stabilized at the IR end of warped throat.
On the other hand,  the visible sector is favored  to be localized
around the UV end of throat in order to realize the high scale
gauge coupling unification at $M_{GUT}\sim 2\times 10^{16}$ GeV.
It turns out that in such setup the
 visible sector soft terms are determined
dominantly by two comparable contributions \cite{mirage1}: the
K\"ahler moduli mediation and the anomaly mediation.
  The resulting soft parameters are unified
at a mirage messenger scale hierarchically lower than $M_{GUT}$,
leading to {\it significantly compressed low energy superparticle
masses} \cite{mirage2,mirageph} compared to other mediation
schemes such as mSUGRA, gauge mediation and anomaly mediation.
Furthermore, under a plausible assumption, mirage mediation
provides more concrete predictions on the superparticle masses,
which have a good chance to be tested at the LHC if the gluino or
squarks are light enough to be copiously produced. In fact, the
two key ingredients of mirage mediation, i.e. (i) brane-localized
SUSY breaking at the IR end of warped geometry  and (ii)
non-perturbative stabilization of the gauge coupling modulus,
might be realized in more generic class of string theories or
brane models \cite{mirage2}.
In this talk, I discuss some features of  SUSY breakdown that occurs
at the tip of throat as in KKLT-type compactification, and also the
low energy superparticle spectrum in the resulting mirage mediation
scheme.

\section{4D effective action of KKLT-type compactification}

One important feature of KKLT-type compactification \cite{kklt} is
the presence of warped throat which is produced  by 3-form fluxes
\cite{gkp}.
The compactified internal space consists of a bulk space which might
be approximately a Calabi-Yau (CY) manifold, and a highly warped
throat attached at CY with SUSY-breaking brane stabilized at its IR
end. In such geometry, the bulk CY  can be identified as the UV end
of throat. To realize the high scale gauge coupling unification, the
visible gauge and matter fields are assumed to live on $D$ branes
stabilized within the bulk CY.

The 4D effective theory of the KKLT-type compactification of Type
IIB string theory includes the UV superfields
$\Phi_{UV}=\{T, U, X\}$ and $V^a,Q^i$, where $T$ and $U$ are the
K\"ahler and complex structure moduli of the bulk CY, $V^a$ and
$Q^i$ are the gauge and matter superfields confined on the visible
sector $D$ branes, and $X$ denotes the open string moduli on those
$D$ branes at the UV side. There are also 4D fields localized at
the IR end of throat, $\Phi_{IR}=\{Z, \Lambda^\alpha\}$, where $Z$
is the throat (complex structure) modulus superfield
parameterizing the size of 3-cycle at the IR end, and
$\Lambda^\alpha$ is the Goldstino superfield confined on
SUSY-breaking brane\footnote{There can be other IR fields, e.g.
the position moduli and gauge fields confined on SUSY-breaking
brane. Those IR fields are not considered here as they do not play
an important role for the transmission of SUSY breakdown.} which
might be an anti-brane as in the original KKLT proposal or any
kind of brane providing SUSY-breaking dynamics.
  In the rigid superspace limit, the Goldstino superfield is
given by \cite{bagger} \bea \Lambda^\alpha=\frac{1}{M_{\rm
SUSY}^2}\xi^\alpha+\theta^\alpha+...,\eea where $\xi^\alpha$ is the
Goldstino fermion,
and  the ellipses stand for the Goldstino-dependent higher order
terms in the $\theta$-expansion. In addition to the above UV and
IR fields, there is of course the 4D SUGRA multiplet which is
quasi-localized in the bulk CY, and also the string dilaton
superfield $S$ whose wavefunction is approximately a constant over
the whole internal space.

The 4D effective action of KKLT-type compactification  takes the
form:\bea \label{4daction1}&& \int d^4x\sqrt{g}\left[\int
d^4\theta \,CC^* \left\{-3\exp \left(-\frac{K}{3}\right)\,\right\}
\right.\nonumber
\\&&+\,\left.\left\{\,\int d^2\theta \,\left( \frac{1}{4}f_a
W^{a\alpha}W^a_\alpha+C^3W\right)+{\rm h.c}\,\right\}\,\right]\eea
where $g_{\mu\nu}$ is the 4D metric in the superconformal frame,
$C=C_0+F^C\theta^2$ is the 4D SUGRA compensator, $K$ is the K\"ahler
 potential, and
 $f_a=T+lS$ ($l=$ rational number)
  are holomorphic gauge kinetic functions which are assumed to
be universal to accommodate the high scale gauge coupling
unification\footnote{Here $\partial f_a/\partial T=1$ can be
considered as our normalization convention of
 $T$.}.
The UV and IR fields are geometrically separated by warped throat,
thus are {\it sequestered} from each other in $e^{-K/3}$:
\bea -3\exp\left(-\frac{K}{3}\right)&=&\Gamma_{UV}+\Gamma_{IR}, \eea
where \bea\Gamma_{UV}&=&
\Gamma_{UV}^{(0)}(S+S^*,\Phi_{UV},\Phi^*_{UV})+{\cal
Y}_i(S+S^*,\Phi_{UV},\Phi^*_{UV})Q^{i*}Q^i, \nonumber
\\\Gamma_{IR}
&=& \Gamma_{IR}^{(0)}(S+S^*,Z,Z^*)+\left(
\frac{C^{*2}}{C}\Lambda^2\Gamma_{IR}^{(1)}(S+S^*,Z,Z^*)+{\rm
h.c}\right)
\nonumber \\
&+&CC^*\Lambda^2\Lambda^{*2}\Gamma_{IR}^{(2)}(S,S^*,Z,Z^*)+..., \eea
where $\Phi_{UV}=\{T, U, X\}$, and  $\Gamma_{IR}$ is expanded in
powers of the Goldstino superfield $\Lambda^\alpha$ and the
superspace derivatives
$D_A=\{\partial_\mu,D_\alpha,\bar{D}_{\dot{\alpha}}\}$.
 The above
effective action is written on flat superspace background and the
SUSY-breaking auxiliary component of the 4D SUGRA multiplet is
encoded in the $F$-component of the compensator $C$. In the
superconformal gauge in which $C=C_0+F^C\theta^2$, the 4D action is
invariant under the rigid Weyl transformation under which \bea
\label{weyl} C\rightarrow e^{-2\sigma}C,\quad
g^C_{\mu\nu}\rightarrow e^{2(\sigma+\sigma^*)}g^C_{\mu\nu},\quad
\theta^\alpha \rightarrow e^{-\sigma+2\sigma^*}\theta^\alpha,\quad
\Lambda^\alpha \rightarrow e^{-\sigma+2\sigma^*}\Lambda^\alpha, \eea
where $\sigma$ is a complex constant, and this determines for
instance the $C$-dependence of $\Gamma_{IR}$.

The effective superpotential of KKLT compactification contains three
pieces: \bea W&=&W_{\rm flux}+W_{\rm np}+W_{\rm Yukawa},\eea where
the flux-induced  $W_{\rm flux}$ stabilizing $S,U,Z,X$ includes the
Gukov-Vafa-Witten superpotential  $W_{GVW}=\int (F_3-4\pi iS
H_3)\wedge \Omega$, where $\Omega$ is the holomorphic $(3,0)$ form
of the underlying CY space,
$W_{\rm np}$ is a non-perturbative superpotential stabilizing $T$,
and finally $W_{\rm Yukawa}$ denotes the Yukawa couplings of the
visible matter fields. Generically, each piece takes the form:
 \bea W_{\rm flux} &=&
\Big({\cal F}(U,X)+\frac{N_{RR}}{2\pi i}Z\ln Z+{\cal O}(Z^2)\Big)
\nonumber \\
&&-\,4\pi
iS\Big({\cal H}(U,X)+N_{NS}Z+{\cal O}(Z^2)\Big),\nonumber \\
W_{\rm np}&=&{\cal A}(U,X)e^{-8\pi^2 (k_1T+l_1S)},
\nonumber \\
W_{\rm Yukawa}&=& \frac{1}{6}\lambda_{ijk}(U,X)Q^iQ^jQ^k,\eea
where $k_1,l_1$ are rational numbers,  $N_{RR},N_{NS}$ are
integers defined as $N_{RR}=\int_{\Sigma} F_3,
N_{NS}=-\int_{\tilde{\Sigma}}H_3$, where $\Sigma$ is the 3-cycle
collapsing  along the throat, $\tilde{\Sigma}$ is its dual
3-cycle, and $F_3$ and $H_3$ are the RR and NS-NS 3-forms,
respectively. Here $Z$ is defined as $\int_{\Sigma}\Omega=Z$, and
then $\int_{\tilde{\Sigma}}\Omega=\frac{1}{2\pi i}Z\ln Z
+\mbox{holomorphic}$ \cite{gkp}. In the above, we assumed that the
axionic shift symmetry of $T$, i.e. $T\rightarrow T+$ imaginary
constant, is preserved by $W_{\rm flux}$ and $W_{\rm Yukawa}$,
while it is broken by $W_{\rm np}$.
To achieve an exponentially small vacuum value of $Z$, which
corresponds to producing a highly warped throat, one needs
$N_{RR}/N_{NS}$ to be positive. The exponential suppression of
$W_{\rm np}$ in the large volume limit ${\rm Re}(T)\gg 1$ implies
that $k_1$ is positive also.

The  above 4D effective action of KKLT-type compactification
involves many model-dependent functions of moduli,
which are difficult to be computed for realistic compactification.
Fortunately, the visible sector soft terms can be determined by only
a few information on the compactification, e.g. the rational
parameters $l,k_1,l_1$ in $f_a$ and $W_{\rm np}$ and the modular
weights which would determine the $T$-dependence of ${\cal Y}_i$,
which can be easily computed or parameterized in a simple manner.
 In particular, soft terms are {\it practically independent} of the detailed forms of
  $\Gamma^{(0)}_{UV}$, $\Gamma_{IR}$, ${\cal F}$, ${\cal H}$,
${\cal A}$ and $\lambda_{ijk}$. This is mainly because (i) the heavy
moduli $\Phi=\{S, U, X\}$ stabilized by flux have negligible
$F$-components, $F^{\Phi}/\Phi \sim m_{3/2}^2/m_{\Phi}\ll
m_{3/2}/8\pi^2$, thus do not participate in SUSY-breaking, and (ii)
the SUSY-breaking IR fields $Z$ and $\Lambda^\alpha$ are sequestered
from the observable sector.

The vacuum value of $Z$ is determined by $W_{\rm flux}$, and
related to the metric warp factor $e^{2A}$ at the tip of throat as
\bea Z \sim \exp \Big(-8\pi^2N_{RR}}S_0/{N_{NS}\Big)\sim e^{3A},\eea
where $S_0$ is the vacuum value of $S$ determined by $D_SW=0$. Since
the scalar component of $CC^*$ corresponds to the conformal factor
of $g^{\mu\nu}$, which can be read off from the Weyl transformation
(\ref{weyl}), $C$ in $\Gamma_{IR}$ should appear in the combination
$Ce^A\sim CZ^{1/3}$. Then the $C$-dependence determined by the Weyl
invariance (\ref{weyl}) suggests \cite{hebecker1} that \bea
\Gamma_{IR}^{(0)}&\sim& (ZZ^*)^{1/3}\,\sim\, e^{2A}, \nonumber \\
\Gamma_{IR}^{(1)}&\sim& Z\,\sim\, e^{3A}, \nonumber \\
\Gamma_{IR}^{(2)}&\sim& (ZZ^*)^{2/3} \,\sim\, e^{4A}\eea for which
\bea m_Z\,\sim\, \frac{F^Z}{Z}\,\sim\, e^A \eea as anticipated. Here
and in the following, unless specified, we use the unit with the 4D
Planck scale $M_{Pl}=1/\sqrt{8\pi G_N}=1$.

The SUSY breaking at the tip of throat provides a positive vacuum
energy density of the order of $M_{\rm SUSY}^4\sim e^{4A}$. This
positive vacuum energy density should be cancelled by the negative
SUGRA contribution of the order of $m_{3/2}^2$, which requires \bea
m_{3/2}\sim e^{2A}.\eea One then finds the following pattern of mass
scales \cite{mirage1}: \bea && m_{S,U,X}\sim
\frac{1}{M_{st}^2R^3}\sim
10^{16} \, {\rm GeV}, \nonumber \\
&& m_Z\sim e^A M_{st}\sim 10^{10} \, {\rm GeV}, \nonumber \\
&&m_{\rm soft}\sim \frac{m_{3/2}}{\ln(M_{Pl}/m_{3/2})}\sim
\frac{m_T}{[\ln(M_{Pl}/m_{3/2})]^2}\sim 10^3\, {\rm GeV},\eea
where $m_{\rm soft}$ denotes the soft masses of the visible
fields, e.g. the gaugino masses, and the string scale $M_{st}$ and
the CY radius $R$ are given by $M_{st}\sim \frac{1}{R}\sim
10^{17}$ GeV.

 The heavy moduli
$S,U,X$ and the throat modulus $Z$ couple to the light visible
fields and $T$ only through the Planck scale suppressed
interactions. Those hidden sector fields can be integrated out to
derive an effective action of $V^a,Q^i,T$ and the Goldstino
superfield $\Lambda^\alpha$ renormalized at a high
scale
near $M_{GUT}$. After
this procedure, the effective action can be written as
\cite{mirage1,mirage2}\bea \label{4daction2}&& \int
d^4x\sqrt{g}\left[\int d^4\theta \, CC^*\Omega_{\rm eff}
+\left\{\,\int d^2\theta \,\left( \frac{1}{4}f^{\rm eff}_a
W^{a\alpha}W^a_\alpha+C^3W_{\rm eff}\right)+{\rm
h.c}\,\right\}\right],\eea where
\bea f_a^{\rm eff}&=&T+lS_0,\nonumber \\
 \Omega_{\rm
eff}&=&-3e^{-K_0/3}+{\cal
Y}_iQ^{i*}Q^i-e^{4A}CC^*\Lambda^2\bar{\Lambda}^2{\cal P}_{\rm
lift}\nonumber
\\
&-&\Big(\frac{e^{3A}C^{*2}}{C}\Lambda^2\Gamma_0+{\rm h.c}\Big),
\nonumber
\\
W_{\rm eff}&=&w_0+{\cal
A}e^{-8\pi^2(k_1T+l_1S_0)}+\frac{1}{6}\lambda_{ijk}Q^iQ^jQ^k,\eea
where $S_0=\langle S\rangle$, $K_0=K_0(T+T^*)$ is the K\"ahler
potential of $T$, $e^{K_0/3}{\cal Y}_i$ is the  K\"ahler metric of
$Q^i$,
${\cal P}_{\rm lift}$ and $\Gamma_0$ are constants of order unity,
and finally $w_0$ is the vacuum value of $W_{\rm flux}$. Note that
at this stage, all of $e^{2A}, {\cal P}_{\rm lift}, \Gamma_0, S_0,
w_0$, and ${\cal A}$ correspond to field-independent constants
obtained after $S,U,X$ and $Z$ are integrated out. As we have
noticed, the condition for vanishing cosmological constant
requires \bea w_0\sim e^{2A}\sim e^{-8\pi^2 l_0 S_0} \qquad
\Big(l_0=\frac{2N_{RR}}{3N_{NS}}>0\Big),\eea and the weak scale
SUSY can be obtained for the warp factor value $e^{2A}\sim
10^{-14}$. For such a small value of  warp factor, one finds that
the vacuum values  of ${\rm Re}(T)$ and the SUSY-breaking
auxiliary components are determined as follows {\it independently
of} the moduli K\"ahler potential $K_0$ \cite{mirage1,mirage2}:
\bea \label{fterms} k_1{\rm Re}(T)&=&(l_0-l_1){\rm Re}(S_0)+{\cal
O}\left(\frac{1}{4\pi^2}\right)\nonumber
\\ \frac{F^C}{C}&=&m_{3/2}\left(1+{\cal
O}\left(\frac{1}{4\pi^2}\right)\right),\nonumber \\
\frac{F^T}{T+T^*}&=&\frac{l_0}{l_0-l_1}\frac{m_{3/2}}{\ln(M_{Pl}/m_{3/2})}\left(1+{\cal
O}\left(\frac{1}{4\pi^2}\right)\right), \nonumber \\
F^{S,U,X}&\sim &\frac{m_{3/2}^2}{m_{S,U,X}}\,\ll
\,\frac{m_{3/2}}{8\pi^2}.\eea Note that  ${\rm Re}(S_0)$, ${\rm
Re}(T)$ and $\frac{1}{g_{GUT}^2}={\rm Re}(T)+l{\rm Re}(S_0)$ are
all required to be positive for $k_1>0$ and $l_0>0$, implying \bea
\label{signs}l_0-l_1\,>\,0,\quad l_0-l_1+k_1l\,>\,0.\eea

One of the interesting features of SUSY breaking at the IR end of
throat is the {\it sequestering} property,
i.e. there is no sizable Goldstino-matter contact term: \bea \Delta
m_i^2CC^*\Lambda^2\bar{\Lambda}^2Q^{i*}Q^i\eea in $\Omega_{\rm eff}$
of (\ref{4daction2}), which would give an additional contribution
$\Delta m_i^2$ to the soft scalar mass-squares. This amounts to that
there is no operator of the form $(ZZ^*)^{1/3}Q^{i*}Q^i$ or
$(ZZ^*)^{2/3}\Lambda^2\bar{\Lambda}^2Q^{i*}Q^i$ in $e^{-K/3}$ of
(\ref{4daction1}). Since $Q^i$ and $\Lambda^\alpha$ are
geometrically separated by warped throat, such contact term can be
generated only by the exchange of bulk field propagating through the
throat. Simple operator analysis assures that the exchange of chiral
multiplet can induce only a higher order operator in the superspace
derivative expansion, while the exchange of light vector multiplet
$\tilde{V}$ can generate the Goldstino-matter contact term with
$\Delta m_i^2\sim \langle D_{\tilde{V}}\rangle$, where
$D_{\tilde{V}}$ is the $D$-component of $\tilde{V}$
\cite{choijeong,hebecker2}. Quite often, throat has an isometry
symmetry providing light vector field which might generate the
Goldtino-matter contact term. However, in many cases, the isometry
vector multiplet does not develop a nonzero $D$-component, and
thereby not generate the contact term
\cite{choijeong,kachrusundrum}. As an example, let us consider the
SUSY breaking by anti-$D3$ brane stabilized at the tip of
Klebanov-Strassler (KS)  throat which has an $SO(4)$ isometry
\cite{ks}. Adding anti-$D3$ at the tip breaks SUSY and also $SO(4)$
down to $SO(3)$. However the unbroken $SO(3)$ assures that the
$SO(4)$ vector multiplets have vanishing $D$-components, thus do not
induce the Goldstino-matter contact term. In fact, this is correct
only up to ignoring the isometry-breaking deformation of KS throat,
which is caused by attaching the throat to  compact CY. Recently,
the effect of such deformation has been estimated
\cite{kachrusundrum}, which found \bea \Delta m_i^2\,\lesssim\,
{\cal O}(e^{\sqrt{28}A})\,\sim\, 10^{-8}m_{3/2}^2.\eea This is small
enough to be ignored compared to the effects of $F^C$ and $F^T$
obtained in (\ref{fterms}).

\section{mirage mediation pattern of soft terms}

The result (\ref{fterms}) on SUSY-breaking $F$-components indicates
that $F^T/T\sim m_{3/2}/4\pi^2 \gg |F^{\Phi}|$ ($\Phi=S,U,X$), and
thus soft terms are determined dominantly by the K\"ahler
moduli-mediated contribution and the one-loop anomaly mediated
contribution which are comparable to each other.
 For the canonically normalized soft terms: \bea
{\cal L}_{\rm soft} &=&
-\frac{1}{2}M_a\lambda^a\lambda^a-\frac{1}{2}m_i^2|\phi^i|^2
-\frac{1}{6}A_{ijk}y_{ijk}\phi^i\phi^j\phi^k+{\rm h.c.}, \eea
where $\lambda^a$ are gauginos, $\phi^i$ are sfermions, $y_{ijk}$
are the canonically normalized Yukawa couplings, the soft
parameters at energy scale just below  $M_{GUT}$ are given by \bea
\label{soft1} M_a &=& M_0 +\frac{b_a}{16\pi^2}g_{GUT}^2m_{3/2},
\nonumber \\
A_{ijk} &=&
\tilde{A}_{ijk}-\frac{1}{16\pi^2}(\gamma_i+\gamma_j+\gamma_k)m_{3/2},
\nonumber \\
m_i^2 &=& \tilde{m}_i^2-\,\frac{1}{32\pi^2}\frac{d\gamma_i}{d\ln
\mu}m_{3/2}^2 \nonumber \\
&+&\frac{1}{4\pi^2}
\left[\sum_{jk}\frac{1}{4}|y_{ijk}|^2\tilde{A}_{ijk} -\sum_a
g^2_aC^a_2(\phi^i)M_0\right]m_{3/2}, \eea where the moduli-mediated
soft masses $M_0$, $\tilde{A}_{ijk}$ and $\tilde{m}_i^2$ are given
by \bea M_0 &=& F^T\partial_T\ln({\rm Re}(f_a)) \nonumber \\
&=& \frac{F^T}{T+T^*}\frac{{\rm Re}(T)}{{\rm Re}(T)+l{\rm
Re}(S_0)}\simeq \frac{F^T}{T+T^*}
\left(\frac{l_0-l_1}{l_0-l_1+k_1l}\right),\nonumber \\
 \tilde{A}_{ijk}&=& F^T\partial_T\ln({\cal Y}_i{\cal Y}_j{\cal Y}_k), \nonumber \\
\tilde{m}^2_i&=& -|F^T|^2\partial_T\partial_{\bar{T}}\ln({\cal
Y}_i), \eea  and $b_a = -3{\rm tr}\left(T_a^2({\rm
Adj})\right)+\sum_i {\rm tr}\left(T^2_a(\phi^i)\right)$, $\gamma_i
=2 \sum_a g_a^2C_2^a(\phi^i)-\frac{1}{2}\sum_{jk}|y_{ijk}|^2$,
where  $C_2^a(\phi^i)=(N^2-1)/2N$ for a fundamental representation
$\phi^i$ of the gauge group $SU(N)$, $C_2^a(\phi^i)=q_i^2$ for the
$U(1)$ charge $q_i$ of $\phi^i$, and
$\omega_{ij}=\sum_{kl}y_{ikl}y^*_{jkl}$ is assumed to be diagonal.

Taking into account the 1-loop RG evolution, the above soft masses
at $M_{GUT}$ lead to the following low energy gaugino masses \bea
M_a(\mu) = M_0 \left[\,1-\frac{1}{8\pi^2}b_ag_a^2(\mu)
\ln\left(\frac{M_{\rm mir}}{\mu}\right)\,\right], \eea showing that
the gaugino masses are unified at the {\it mirage messsenger scale}
\cite{mirage2}: \bea \label{mirage-scale} M_{\rm mir} =
\frac{M_{GUT}}{(M_{Pl}/m_{3/2})^{\alpha/2}}, \eea where \bea
\label{alpha} \alpha \equiv \frac{m_{3/2}}{M_0\ln(M_{Pl}/m_{3/2})} =
\frac{l_0-l_1+k_1l}{l_0} \left(1+{\cal
O}\left(\frac{1}{4\pi^2}\right)\right),\eea
 while the gauge
couplings are still unified at $M_{GUT}=2\times 10^{16}$ GeV. The
low energy values of $A_{ijk}$ and $m_i^2$ generically depend on the
associated Yukawa couplings $y_{ijk}$. However if $y_{ijk}$ are
negligible or if
$\tilde{A}_{ijk}/M_0=(\tilde{m}^2_i+\tilde{m}^2_j+\tilde{m}^2_k)/M_0^2=1$,
their low energy values also show  the mirage unification feature
\cite{mirage2}:\bea \label{mirage-solution} A_{ijk}(\mu) &=&
\tilde{A}_{ijk}+ \frac{M_0}{8\pi^2}(\gamma_i(\mu) +
\gamma_j(\mu)+\gamma_k(\mu)) \ln\left(\frac{M_{\rm
mir}}{\mu}\right),
\nonumber \\
m_i^2(\mu) &=& \tilde{m}_i^2-\frac{M^2_0}{8\pi^2}Y_i\left(
\sum_jc_jY_j\right)g^2_Y(\mu)\ln\left(\frac{M_{GUT}}{\mu}\right)
\nonumber \\
&& \hspace{1cm} +\,\frac{M_0^2}{4\pi^2}\left\{
\gamma_i(\mu)-\frac{1}{2}\frac{d\gamma_i(\mu)}{d\ln \mu}\ln\left(
\frac{M_{\rm mir}}{\mu}\right)\right\} \ln\left( \frac{M_{\rm
mir}}{\mu}\right), \eea where $Y_i$ is the $U(1)_Y$ charge of
$\phi^i$. Quite often, the moduli-mediated squark and slepton masses
have a common value, i.e.
$\tilde{m}^2_{\tilde{Q}}=\tilde{m}^2_{\tilde{L}}$, and then the
 squark and slepton masses of the  1st and 2nd generation are unified again
at $M_{\rm mir}$.

In regard to phenomenology,  the most interesting feature of
mirage mediation is that it gives rise to {\it significantly
compressed low energy SUSY spectrum} compared to other popular
schemes such as mSUGRA, gauge mediation and anomaly mediation.
This feature can be easily understood by noting that soft
parameters are unified at $M_{\rm
mir}=M_{GUT}(m_{3/2}/M_{Pl})^{\alpha/2}$ which is hierarchically
lower than $M_{GUT}$  if $\alpha$ has  a positive value of order
unity.  Indeed, the result (\ref{alpha}) shows  that $\alpha$ is
(approximately) a positive rational number for the rational
numbers $k_1,l,l_0,l_1$ obeying the constraints (\ref{signs}).
Another, but related, interesting feature of mirage mediation is
that the little SUSY fine tuning problem of the MSSM can be
significantly ameliorated in TeV scale mirage mediation scenario
with $M_{\rm mir}\sim 1$ TeV, i.e. $\alpha\simeq 2$
\cite{mirage2,little}.

In fact, mirage mediation provides more concrete prediction under
a rather plausible assumption. Assuming that $f_a$ are
(approximately) universal, which might be required to realize the
gauge coupling unification at $M_{GUT}$, the low scale gaugino
masses at the RG point $\mu\sim 500$ GeV are given by
 \bea M_1&\simeq & M_0(0.42+0.28\alpha),\nonumber \\
M_2&\simeq & M_0(0.83+0.085\alpha),\nonumber \\
M_3&\simeq& M_0(2.5-0.76\alpha),\eea leading to \cite{choinilles}
\bea M_1:M_2:M_3\,\simeq\,
(1+0.66\alpha):(2+0.2\alpha):(6-1.8\alpha).\eea
 The low scale masses  of
the 1st and 2nd generations of squarks and sleptons are also
easily obtained to be \bea m_{\tilde{Q}}^2&\simeq&
\tilde{m}^2_{\tilde{Q}}+M_0^2(5.0-3.6\alpha+0.51\alpha^2),\nonumber
\\
m_{\tilde{D}}^2&\simeq&\tilde{m}^2_{\tilde{D}}+M_0^2(4.5-3.3\alpha+0.52\alpha^2),\nonumber
\\
m_{\tilde{L}}^2&\simeq&\tilde{m}^2_{\tilde{L}}+M_0^2(0.49-0.23\alpha-0.015\alpha^2),\nonumber
\\
m_{\tilde{E}}^2&\simeq&\tilde{m}^2_{\tilde{E}}+M_0^2(0.15-0.046\alpha-0.016\alpha^2),
\eea where $\tilde{Q},\tilde{D},\tilde{L}$ and $\tilde{E}$ denote
the $SU(2)_L$ doublet squark, singlet up-squark, singlet
down-squark, doublet lepton, and singlet lepton, respectively.
Assuming that the matter K\"ahler metrics obey simple unification
(or universality) relations such as ${\cal Y}_Q={\cal Y}_E$ and
${\cal Y}_D={\cal Y}_L$ which would yield
$\tilde{m}^2_{\tilde{Q}}= \tilde{m}^2_{\tilde{E}}$ and
$\tilde{m}^2_{\tilde{D}}=\tilde{m}^2_{\tilde{L}}$, we find \bea
&&M_1^2:(m_{\tilde{Q}}^2-m_{\tilde{E}}^2):(m_{\tilde{D}}^2-m_{\tilde{L}}^2)
\nonumber \\&\simeq&  (0.18+0.24\alpha+0.09\alpha^2)
:(4.9-3.5\alpha+0.53 \alpha^2):(4.0-3.1\alpha+0.54\alpha^2).\eea
Note that these ratios are independent of the presence of extra
matter fields at scales above TeV.

If the idea of low energy SUSY is correct and the gluino or squark
masses are lighter than 2 TeV,  some superparticle masses, e.g.
the gluino mass and the first two neutralino masses as well as
some of the squark and slepton masses, might be determined at the
LHC by analyzing various kinematic invariants of the cascade
decays of gluinos and squarks. It is then quite probable that the
LHC measurements of those superparticle masses are good enough to
test the above predictions of mirage mediation \cite{cho}.

\section{conclusion}
Warped throat appears often in  fluxed compactification of string
theory. If  SUSY-breaking brane carrying a positive energy density
is introduced into the compactification geometry containing warped
throat, it is naturally stabilized at the  tip of throat. On the
other hand, the high scale gauge coupling unification at
$M_{GUT}\sim 2\times 10^{16}$ GeV suggests that the visible gauge
and matter fields are localized in the bulk space corresponding to
the UV end of throat. If (some of) the  moduli  which determine
the 4D gauge couplings were stabilized (before introducing
SUSY-breaking brane)
by non-perturbative dynamics at a SUSY-preserving configuration as
in the KKLT compactification, the SUSY-breaking brane at the tip
of throat leads to a highly distinctive pattern of soft terms of
the visible fields localized at the UV end of throat. The
resulting soft parameters are unified at a mirage messenger scale
hierarchically lower than $M_{GUT}$, while the gauge couplings are
unified still at $M_{GUT}$, leading to  the term ``mirage
mediation". The low energy superparticle masses in mirage
mediation are significantly compressed compared to those in
mSUGRA, gauge mediation and anomaly mediation. Furthermore, under
a plausible assumption, the scheme provides more concrete
predictions on the superparticle masses, which might be tested at
the LHC.

\vspace{5mm} \noindent{\large\bf Acknowledgments} \vspace{5mm}

This work is supported by the KRF Grant funded by the Korean
Government (KRF-2005-201-C00006), the KOSEF Grant (KOSEF
R01-2005-000-10404-0), and the Center for High Energy Physics of
Kyungpook National University. I thank  W. S. Cho, K. S. Jeong, Y.
G. Kim, T. Kobayashi, H. P. Nilles, and K. Okumura for
collaborations and useful discussions.

\end{document}